\input{aipcheck}
\documentclass[final]{aipproc}
\layoutstyle{6x9}

\begin{document}

\title{Antiferromagnets at Low Temperatures}


\author{C.\ P.\ Hofmann}
{address={Facultad de Ciencias, Universidad de Colima,
Bernal D\'iaz del Castillo 340, Colima C.P.\ 28045, Mexico}}

\begin{abstract}
The low-temperature properties of the Heisenberg antiferromagnet in 2+1 space-time dimensions are analyzed within the framework of
effective Lagrangians. It is shown that the magnon-magnon interaction is very weak and repulsive, manifesting itself through a term
proportional to five powers of the temperature in the pressure. The structure of the low-temperature series for antiferromagnets in
2+1 dimensions is compared with the structure of the analogous series for antiferromagnets in 3+1 dimensions. The model-independent
and systematic effective field theory approach clearly proves to be superior to conventional condensed matter methods such as
spin-wave theory.
\end{abstract}

\maketitle

\section{Motivation}

In the present study we focus our attention on the low-energy properties of antiferromagnets in dimension $d$=2+1. A thorough
analysis of these condensed matter systems, using effective field theory methods, was performed in
Refs.~\citep{HL90,CHN89,NZ89,F89,HN93}.
Here, we go one step further in the perturbative expansion, taking into account contributions to the free energy density up to
three-loop order. The improvement by going from two- to three-loop order is that the interaction among the spin-wave degrees of
freedom only starts manifesting itself at the three-loop level. We show that the magnon-magnon interaction in the O(3)
antiferromagnet in $d$=2+1 -- the O(3)-invariant quantum Heisenberg antiferromagnet on a square or a honeycomb lattice -- is
very weak and repulsive and manifests itself through a term proportional to five powers of the temperature in the free energy 
density.

Although our analysis is general, referring to any system which exhibits a spontaneously broken symmetry O($N$) $\to$ O($N$-1) and
a Lorentz-invariant leading-order effective Lagrangian, our interest will be devoted to the special case $N$=3. Here the
internal O(3) spin symmetry of the isotropic Heisenberg model is spontaneously broken by the ground state which displays a non-zero
staggered magnetization.

One may wonder why the quantum Heisenberg antiferromagnet,
\begin{equation}
\label{HeisenbergModel}
{\cal H} = -J \sum_{n.n.} {\vec S}_m \cdot {\vec S}_n \, , \qquad J=const. \, , 
\end{equation}
represents a system described by a Lorentz-invariant leading-order effective Lagrangian. After all, the lattice structure of a
solid singles out preferred directions, such that the effective Lagrangian in general is not even invariant under space rotations.
In the case of a square lattice, however, the anisotropy only shows up at higher orders of the derivative expansion
\citep{HN93} -- the discrete symmetries of the two-dimensional system thus imply space rotation symmetry
at leading order in the effective expansion. The same is true for an antiferromagnet defined on a honeycomb lattice. Hence, the
leading-order effective Lagrangian describing the quantum Heisenberg antiferromagnet on a square or honeycomb lattice is invariant
under space rotations and can be brought to a (pseudo-) Lorentz-invariant form \citep{L94a}: Antiferromagnetic spin-wave
excitations exhibit relativistic kinematics, with the velocity of light replaced by the spin-wave velocity.

\section{Effective field theory evaluation}

In a Lorentz-invariant framework the construction of effective Lagrangians is straightforward \citep{L94b}: One
writes down the most general expression consistent with Lorentz symmetry and the internal, spontaneously broken symmetry G of the
underlying model in terms of Goldstone fields $U^a(x), a = 1, \dots$, dim(G)-dim(H) -- the effective Lagrangian then consists of a
string of terms involving an increasing number of derivatives or, equivalently, amounts to an expansion in powers of the momentum.

In the particular case we are considering, the symmetry G = O($N$) is explicitly broken by an external field. It is convenient to
collect the ($N$-1) Goldstone fields $U^a$ in a $N$-dimensional vector $U^i = (U^0,U^a)$ of unit length,
\begin{equation}
U^i(x) \, U^i(x) \, = \, 1 \, ,
\end{equation}
and to take the constant external field along the zeroth axis, $H^i = (H,0, \dots , 0)$. The Euclidean form of the effective
Lagrangian up to and including order $p^4$ then reads \citep{HL90}:
\begin{eqnarray}
\label{Leff}
{\cal L}_{eff} & = & {\cal L}^2_{eff} + {\cal L}^4_{eff} \, , \nonumber \\
{\cal L}^2_{eff} & = &  \mbox{$ \frac{1}{2}$} F^2 {\partial}_{\mu}
U^i{\partial}_{\mu} U^i \, - \, {\Sigma}_s H^i U^i  \, , \nonumber\\
{\cal L}^4_{eff} & = & \, - \, e_1 ({\partial}_{\mu} U^i
{\partial}_{\mu} U^i)^2 \, - \, e_2 \, ({\partial}_{\mu} U^i {\partial}_{\nu}
U^i)^2 \nonumber \\
& & + \, k_1 \! \, \frac{{\Sigma}_s}{F^2} \, (H^i U^i)
({\partial}_{\mu} U^k {\partial}_{\mu} U^k) \,
- k_2 \, \! \frac{{\Sigma}_s^2}{F^4} \, (H^i U^i)^2 \,
- \, k_3 \, \! \frac{{\Sigma}_s^2}{F^4} \, H^i \! H^i  \, . \nonumber
\end{eqnarray}

In the effective Lagrangian framework at finite temperature, the partition function is represented as a Euclidean functional
integral
\begin{equation}
\label{TempExp}
\mbox{Tr} \, [\exp(-{\cal H}/T)] \, = \, \int [{\mbox{d}}U] \,
\exp \Big( - {\int}_{\!\!\! {\cal T}} \! {\mbox{d}}^4x \, {\cal L}_{eff} \Big)
\, , \nonumber
\end{equation}
where the integration is performed over all field configurations which are periodic in the Euclidean time direction:
$U({\vec x}, x_4 + \beta) = U({\vec x}, x_4)$, with $\beta \equiv 1/T$.

\begin{figure}
\label{figure1}
\resizebox{.64\columnwidth}{!}
{\includegraphics{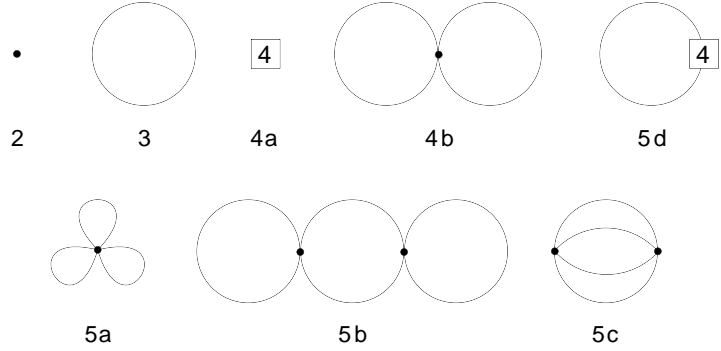}}
\caption{\it Feynman graphs related to the low-temperature expansion of the
partition function for an O($N$) antiferromagnet up to three-loop order in
dimension $d$=2+1. The numbers attached to the vertices refer to the piece of
the effective Lagrangian they come from. Vertices associated with the leading
term ${\cal L}^2_{eff}$ are denoted by a dot. Note that loops are suppressed
by one momentum power in $d$=2+1.}
\end{figure}

We have evaluated the partition function of an O($N$) antiferromagnet in dimension $d$=2+1 up to three-loop order -- the relevant
Feynman graphs are shown in Fig.\ref{figure1}. 

One notices that the spatial anisotropies which indeed start manifesting themselves at next-to-leading order in the effective
Lagrangian through terms like
\begin{equation}
\label{extraTerm1}
\sum_{s=1,2} \, {\partial}_s {\partial}_s U^i \, {\partial}_s {\partial}_s U^i \, , \nonumber
\end{equation}
cannot manifest themselves in the magnon-magnon interaction up to the order $p^5$ considered in the present work: although they
give rise to an additional term in the free energy density involving five powers of the temperature, this is a purely kinematical
effect related to the one-loop graph 5d. 

\section{Results}

We are particularly interested in the limit $T \gg M_{\pi}$ which we implement by holding $T$ fixed and sending $M_{\pi}$ (or,
equivalently, the external field $H$) to zero. The pressure $P$, the energy density $u$, the entropy density $s$,
and the heat capacity $c_V$ for the O(3) antiferromagnet in 2+1 dimensions are then given by
\begin{eqnarray}
P & = & \frac{\zeta(3)}{\pi} \, T^3 \, \Big[ 1 - \frac{\pi q_1}{\zeta(3)} \,
\frac{T^2}{F^4} + {\cal O}(T^3) \Big]
\approx 0.3826 \, T^3 \,
\Big[ 1 + 0.02294 \, \frac{T^2}{F^4} + {\cal O}(T^3) \Big] \, ,
\nonumber \\
u & = & \frac{2 \, \zeta(3)}{\pi} \, T^3 \, \Big[ 1 -
\frac{2 \pi q_1}{\zeta(3)} \, \frac{T^2}{F^4} + {\cal O}(T^3) \Big]
\approx 0.7653 \, T^3 \, \Big[ 1 + 0.04589 \, \frac{T^2}{F^4}
+ {\cal O}(T^3) \Big] \, ,
\nonumber \\
s & = & \frac{3 \,\zeta(3)}{\pi} \, T^2 \, \Big[ 1 -
\frac{5 \pi q_1}{3 \zeta(3)} \, \frac{T^2}{F^4} + {\cal O}(T^3) \Big]
\approx 1.1479 \, T^2 \, \Big[ 1 + 0.03824 \, \frac{T^2}{F^4}
+ {\cal O}(T^3)\Big] \, ,
\nonumber \\
c_V & = & \frac{6 \, \zeta(3)}{\pi} \, T^2 \, \Big[1 -
\frac{10 \pi q_1}{3 \zeta(3)} \, \frac{T^2}{F^4} + {\cal O}(T^3) \Big]
\approx 2.2958 \, T^2 \Big[ 1 + 0.07648 \, \frac{T^2}{F^4}
+ {\cal O}(T^3) \Big] \, .
\end{eqnarray}
The coefficient $q_1$ is a pure number, originating from the numerical evaluation of the three-loop graph 5c.
The respective first terms in the above series represent the free Bose gas contribution which originates from a one-loop graph. The
effective interaction among the Goldstone bosons only manifests itself through a term of order $T^5$ in the pressure, related to a
three-loop graph. Interestingly, the coefficient $q_1$ is negative, such that the magnon-magnon-interaction in the O(3)
antiferromagnet in $d$=2+1 is repulsive at low temperatures. It is remarkable that the coefficient of the interaction term in these
series is fully determined by the symmetries inherent in the leading-order effective Lagrangian, and does not involve any
next-to-leading order coupling constants from ${\cal L}^4_{eff}$, reflecting the anisotropies of the square lattice or the
Lorentz-noninvariant nature of the quantum Heisenberg antiferromagnet defined on a square or a honeycomb lattice -- the symmetry is
thus very restrictive in $d$=2+1.

The fact that an interaction term proportional to four powers of the temperature does not show up in the temperature expansion for
the pressure of the O(3) antiferromagnet in the limit $T \gg M_{\pi}$, was already pointed out in Ref.~\citep{HN93}:
this was an effective Lagrangian calculation that operated on the two-loop level. We are not aware of any microscopic calculation
that aimed at this accuracy. Moreover, our result that the leading contribution of the magnon-magnon interaction in the pressure is
repulsive and of order $T^5$ requires a three-loop calculation on the effective level, performed in the present study - it is
probably fair to say that this accuracy is beyond the reach of any realistic microscopic calculation based on spin-wave theory.

We now want to compare the low-temperature series for antiferromagnets in 2+1 dimensions with those for antiferromagnets in 3+1
dimensions, which take the form
\begin{eqnarray}
P & = & \mbox{$ \frac{1}{45}$} {\pi}^2 \, T^4 \Bigg[ 1 \, + \,
\frac{1}{72} \, \frac{T^4}{F^4} \, \ln{\frac{T_p}{T}} \,
+ \, {\cal O} (T^6) \, \Bigg] \, , \nonumber \\
u & = & \mbox{$ \frac{1}{15}$} {\pi}^2 \, T^4 \Bigg[ 1 \, + \,
\frac{1}{216} \, \frac{T^4}{F^4} \Big(7 \,
\ln{\frac{T_p}{T}} - 1 \Big) \, + \, {\cal O} (T^6) \, \Bigg]
\, , \nonumber \\
s & = & \mbox{$ \frac{4}{45}$} {\pi}^2 \, T^3 \Bigg[ 1 \, + \,
\frac{1}{288} \, \frac{T^4}{F^4} \Big(8 \, \ln \frac{T_p}{T}
- 1\Big) \, + \, {\cal O} (T^6) \, \Bigg] \, , \nonumber \\
c_V & = & \mbox{$ \frac{4}{15}$} {\pi}^2 \, T^3 \Bigg[ 1 \, +
\, \frac{1}{864} \, \frac{T^4}{F^4} \Big(56 \,
\ln{\frac{T_p}{T}} - 15 \Big) \, + \, {\cal O} (T^6) \, \Bigg] \, .
\end{eqnarray} 
The respective first contributions represent the free Bose gas term which originates from a one-loop graph, whereas the effective
interaction among the Goldstone bosons, remarkably, only manifests itself through a term of order $T^8$ in the pressure. This
contribution contains a logarithm, characteristic of the effective Lagrangian method in four space-time dimensions, which involves a
scale, $T_p$, involving coupling constants from ${\cal L}^4_{eff}$. At low temperatures, the logarithm $\ln[T_p/T]$ in the pressure
is positive, such that the interaction among the Goldstone bosons in $d$=3+1, in the absence of an external field $H$, is repulsive,
much like in $d$=2+1. The symmetries in $d$=3+1, however, are somewhat less restrictive than in $d$=2+1, where the interaction term is
unambiguously determined by the coupling constant $F$ of the leading order effective Lagrangian. Details of the calculation and a
more elaborate discussion of the results can be found in the original articles \citep{H99,H10}.

\section{Conclusions}

Our main result is that the interaction among magnons in the O(3)-invariant Heisenberg antiferromagnet, defined on a square or a
honeycomb lattice, is very weak and repulsive at low temperatures, manifesting itself through a term proportional to five powers of
the temperature in the pressure. Remarkably, the coefficient of this interaction term is fully determined by the leading-order
effective Lagrangian ${\cal L}^2_{eff}$ and does not involve any higher order effective constants from ${\cal L}^4_{eff}$. Additional
effective constants in ${\cal L}^4_{eff}$, taking into account the Lorentz-noninvariant nature of the system, merely affect the
renormalization of the magnon mass or yield higher-order corrections to the magnon dispersion law, but do not affect at all the
leading contribution originating from the magnon-magnon interaction in the pressure.

We would like to emphasize that the order of the calculation presented here, appears to be beyond the reach of any realistic
microscopic calculation based on spin-wave theory or other standard condensed matter methods methods, such as Schwinger boson mean
field theory. The fully systematic effective Lagrangian method thus clearly proves to be more efficient than the complicated
microscopic analysis. Another virtue of the effective Lagrangian technique is that it addresses the problem from a unified and
model-independent point of view based on symmetry -- at large wavelengths, the microscopic structure of the system only manifests
itself in the numerical values of a few coupling constants.

\begin{theacknowledgments}
Support by CONACYT grant No. 50744-F is gratefully acknowledged.
\end{theacknowledgments}

\end{document}